\documentclass[hyper]{prop2015}
\usepackage[english]{babel}
\usepackage{bbm}
\usepackage{mathrsfs}
\usepackage{tikz}
\usetikzlibrary{matrix,cd,arrows}

\newcommand{\der}[1]{\frac{\partial}{\partial #1}}

\newcommand{\eand}{{~~~\mbox{and}~~~}}

\newcommand{\dd}{\mathrm{d}}
\def\slasha#1{\setbox0=\hbox{$#1$}#1\hskip-\wd0\hbox to\wd0{\hss\sl/\/\hss}}

\newcommand{\im}{\mathrm{im}}

\newcommand{\frg}{\mathfrak{g}}
\newcommand{\frh}{\mathfrak{h}}
\newcommand{\FR}{\mathbbm{R}}     

\newcommand{\ce}{\mathsf{CE}}
\newcommand{\weil}{\mathsf{W}}
\newcommand{\inv}{\mathsf{inv}}
\newcommand{\free}{\mathsf{F}}

\newcommand{\sG}{\mathsf{G}}

\newcommand{\NN}{\mathbbm{N}}     			

\newcommand{\CF}{\mathcal{F}}
\newcommand{\CH}{\mathcal{H}}
\newcommand{\CG}{\mathcal{G}}

\newcommand{\dder}[1]{\frac{\dd}{\dd #1}}   		

\newcommand{\astringsk}{\mathfrak{string}_{\rm sk}}
\newcommand{\astringskext}{\mathfrak{string}_{\rm sk}^{\rm ext}}
\newcommand{\twastringsk}{\widehat{\mathfrak{string}}_{\rm sk}}
\newcommand{\twastringl}{\widehat{\mathfrak{string}}_{\hat \Omega}}
\newcommand{\astringl}{\mathfrak{string}_{\hat \Omega}}

\newcommand{\RZ}{\mathbbm{Z}}     			

\newcommand{\CCC}{\mathscr{C}}

\newcommand{\CD}{\mathcal{D}}

\category{Proceedings}
\keywords{Higher gauge theory, $L_\infty$-algebras, string Lie 2-algebra, fake flatness}
\subtitle{\href{http://www.maths.dur.ac.uk/lms/109/index.html}{LMS/EPSRC Durham Symposium on Higher Structures in M-Theory}}
\title{Twisted Weil Algebras for the String Lie 2-Algebra}

\begin{acknowledgements}
I would like to thank my supervisor Christian S\"amann for his guidance and for working with me on the papers~\cite{Saemann:2017rjm,Saemann:2017zpd}, which are the basis of this review.  I gratefully acknowledge being supported in part by STFC award 1669147 and in part by MOE tier 2 grant R-144-000-396-112. Further thanks go to the Glasgow Mathematical Journal Trust for their support during the LMS/EPSRC Durham Symposium.
\end{acknowledgements}

\author[Lennart Schmidt]{Lennart Schmidt\inst{a,}\footnote{Corresponding author e-mail:~\href{mailto:phylen@nus.edu.sg}{\textsf{phylen@nus.edu.sg}}}}
\address[1]{Department of Physics, National University of Singapore, 2 Science Drive 3, Singapore 117551}
\shortauthors{L.~Schmidt}
\begin{abstract}
In this article, we give a concise summary of $L_\infty$-algebras viewed in terms of Chevalley--Eilenberg algebras, Weil algebras and invariant polynomials and their use in defining connections in higher gauge theory. Using this, we discuss the example of the string Lie 2-algebra in both the skeletal and the loop model. In both cases, we show how to arrive at the twisted Weil algebras which were used in~\cite{Saemann:2017rjm} to construct a non-abelian self-dual string soliton, see also~\cite{Saemann:2017zpd,Sati:2008eg,Sati:2009ic}.
\end{abstract}
\shortabstract
\begin{document}
\maketitle

\section{Introduction}
Higher gauge theory describes the parallel transport of extended objects~\cite{Baez:2010ya,Baez:2002jn} and is therefore of central interest to string and M-theory whose fundamental objects are higher dimensional. The parallel transport is described by higher connections taking values in $L_\infty$-algebras, just as ordinary gauge theory employs connections taking values in Lie algebras. One of the challenges in finding interesting examples of higher gauge theories is to choose the right gauge structure, that is, to choose an appropriate $L_\infty$-algebra. In~\cite{Saemann:2017rjm}, we used a twisted version of the string Lie 2-algebra, which is based on a modified Weil algebra, to construct a non-abelian self-dual string soliton, which is the M-theory analogue of the 't Hooft--Polyakov monopole. 

The aim of this article is to explain this twisted Weil algebra for the string Lie 2-algebra and show how one can arrive at the twist. Here, the string Lie 2-algebra appears in two different guises: the skeletal model and the loop model. The twist in the skeletal model already appeared and was derived in~\cite{Sati:2008eg,Fiorenza:2010mh,Sati:2009ic}. In~\cite{Saemann:2017rjm}, this was extended to the loop model. The twisted Weil algebra of the skeletal model also frequently appears in the context of and is closely related to string structures, see~\cite{Waldorf:2009uf,Stolz:2004aa,Carey:0410013}. It is, e.g., compatible with the gauge structure of the (1,0)-models introduced in~\cite{Samtleben:2011fj}, a fact, which was used in~\cite{Saemann:2017zpd} to improve on the model.

We start by giving a concise review of $L_\infty$-algebras viewed as differential graded commutative algebras, which leads us to discuss the Chevalley--Eilenberg algebra, the Weil algebra and the invariant polynomials associated with an $L_\infty$-algebra. We discuss morphisms between $L_\infty$-algebras and, as this is crucial to the calculation of the twists, introduce an explicit definition of 2-morphisms of $L_\infty$-algebras due to~\cite{Sati:2008eg}. As examples, we discuss the two models of the string Lie 2-algebra in a little more detail. 

We then introduce higher gauge theory from morphisms of differential graded commutative algebras --- an approach in its full extent due to~\cite{Sati:2008eg}. Using the example of the string Lie 2-algebra, we briefly comment on the fake flatness condition, which frequently appears in higher gauge theory, and the need to move to twisted Weil algebras to allow for this condition to be relaxed. Lastly, we use the above concepts to derive these modified Weil algebras for both the skeletal and the loop model.

\section{\texorpdfstring{$L_\infty$}{L_infinity}-algebras in their various guises}
Many of the higher structures arising in string and M-theory appear in the form of categories and 2-categories. A \textit{category} $\CCC=\CCC_0\leftleftarrows \CCC_1$ consists out of a collection of objects $\CCC_0$ and a collection of morphisms $\CCC_1$ between those objects. Furthermore, these morphisms can be composed associatively and there is an identity morphism associated to each object. In its simplest form~\cite{Lack:2010sj}, a \textit{2-category} is additionally equipped with a collection of 2-morphisms between morphisms, that can be composed in two ways --- vertically and horizontally. Both these compositions are associative, have appropriate identity 2-morphisms and behave well with respect to each other, see e.g.~\cite{Baez:2010ya} for more details. 

Here, the important aspect of 2-morphisms is that they allow us to describe categorically equivalent objects, just as homotopies allow to define homotopy equivalent spaces. More precisely, we consider two given objects $x$ and $y$ to be equivalent if there are morphisms $f:x\to y$ and $g:y\to x$ such that their compositions $f\circ g$ and $g\circ f$ are connected to the relevant identity morphisms via a 2-morphism. It is this notion that will play a central role in the later discussion.

In higher gauge theory, the most relevant higher structure is that of an \textit{$L_\infty$-algebra}. These can be seen as categorified versions of an ordinary Lie algebra~\cite{Baez:2003aa}. In their most familiar guise, they consist out of a $\RZ$-graded vector space $\frg = \sum_{k\in \RZ} \frg_k$ together with a set of totally antisymmetric, multilinear structure maps \mbox{$\mu_i : \wedge^i \frg \to \frg,~i\in \NN,$} of degree $2-i$, which satisfy the \textit{higher Jacobi relations}
\begin{equation}\label{eq:hom_rel}
\begin{aligned}
&\sum\limits_{i+j=n}\,\, \sum\limits_{\sigma\in S_{i|j}} (-1)^{ij} \chi(\sigma; l_1,\dots,l_n)\times\\
&\kern.5cm\times\mu_{j+1}(\mu_i(l_{\sigma(1)},\dots,l_{\sigma(i)}),l_{\sigma(i+1)},\dots,l_{\sigma(n)})= 0
\end{aligned}
\end{equation}
for all $n\in\NN$ and $l_1,\dots,l_n\in \frg$, where the second sum runs over all $(i,j)$-\textit{unshuffles} $\sigma\in S_{i|j}$, that is, permutations whose image consists of ordered sets of length $i$ and $j$: $\sigma(k)<\sigma(k+1)$ for $k\neq i$. Additionally, $\chi(\sigma; l_1,\dots,l_n)$ denotes the \textit{graded antisymmetric Koszul sign} defined by the graded antisymmetrized products
\begin{equation}
l_1\wedge\ldots \wedge l_n = \chi(\sigma;l_1,\ldots,l_n)l_{\sigma(1)}\wedge\ldots\wedge l_{\sigma(n)}~,
\end{equation}
where any transposition not involving only odd degree elements acquires a minus sign. An \textit{$n$-term $L_\infty$-algebra} is an $L_\infty$-algebra that is concentrated (i.e.~non-trivial only) in degrees $-n+1,\dots,0$. A 1-term $L_\infty$-algebra then corresponds to an ordinary Lie algebra. For more details and references see also~\cite{contrib:saemann}.

An alternative and elegant way of describing $L_\infty$-algebras and their higher Jacobi relations is given in terms of coalgebras and coderivations, see~\cite{Lada:1992wc} and\linebreak also~\cite{Lada:1994mn}. To see this, consider an $n$-term $L_\infty$-algebra $\frg\left[-1\right]$, where we shift the degree of all elements by $-1$. This consequently shifts the degree of the maps $\mu_i$ from $2-i$ to $1$ and allows to define a degree $1$ coderivation
\begin{equation}
\CD:\vee^\bullet \frg[-1]\to\vee^\bullet \frg[-1]~,~~~\CD^1=\sum\limits_i \mu_i~,
\end{equation}
which acts on the graded symmetric coalgebra $\vee^\bullet \frg[-1]$ generated by $\frg[-1].$ More explicitly, $\vee^\bullet \frg[-1]$ is spanned by graded symmetric elements $l_1 \vee\dots\vee\, l_n$ and is equipped with the coproduct 
\pagebreak
\begin{equation}
\begin{aligned}
&\Delta(l_1\vee\dots\vee l_n) = \sum\limits_{i+j=n}\,\,\sum\limits_{\sigma\in S_{i|j}} \epsilon(\sigma;l_1,\dots,l_n)\times\\
&\kern2cm\times (l_{\sigma(1)}\vee\dots\vee l_{\sigma(i)})\otimes(l_{\sigma(i+1)}\vee\dots\vee l_{\sigma(n)})~,
\end{aligned}
\end{equation}
where $S_{i|j}$ again denotes the set of $(i,j)$-unshuffles and $\epsilon$ is now the \textit{graded symmetric Koszul sign}, which is related to the graded antisymmetric Koszul sign via the equation
\begin{equation}
\epsilon(\sigma;l_1,\dots,l_n) = \text{sgn}(\sigma)\chi(\sigma;l_1,\dots,l_n)~.
\end{equation}
A coderivation $\CD$ is now given by a linear map $\CD:\vee^\bullet \frg[-1]\to\vee^\bullet\frg[-1]$ which satisfies the co-Leibniz rule
\begin{equation}
\Delta\circ \CD = (\CD \otimes \text{id}+\text{id} \otimes \CD) \circ \Delta~.
\end{equation}
Such a coderivation $\CD$ is completely determined by $\CD^1$, that is, its image restricted to $\frg[-1]$, as the co-Leibniz rule implies~\cite{0387950680,Fregier:2013dda}
\begin{equation}
\begin{aligned}
\CD(l_1\vee&\dots\vee l_n) =\sum\limits_{i+j=n}\,\sum\limits_{\sigma\in S_{i|j}} \epsilon(\sigma;l_1,\dots,l_n)\times\\
& \times\CD^1(l_{\sigma(1)}\vee\dots\vee l_{\sigma(i)})\vee l_{\sigma(i+1)}\vee\dots\vee l_{\sigma(n)}~.
\end{aligned}
\end{equation}
With $\CD^1$ given by $\sum \mu_i$ as above, the condition $\CD^2=0$ then exactly corresponds to the higher homotopy relations~\eqref{eq:hom_rel} and, thus, coalgebras yield an equivalent way of describing $n$-term $L_\infty$-algebras.

Here, however, we will adopt a third way of describing $n$-term $L_\infty$-algebras which arises when dualizing the above concept\footnote{Care must be taken in the infinite-dimensional case.}. The graded symmetric coalgebra $\vee^\bullet\frg[-1]$ becomes the graded symmetric algebra $\wedge^\bullet \frg^\ast[1]$, where, in the process of dualizing all gradings acquire a minus sign. The coderivation $\CD$ then induces a degree $1$ differential $Q:\wedge^\bullet \frg^\ast[1]\to\wedge^\bullet \frg^\ast[1]$, acting on the graded symmetric algebra $\wedge^\bullet \frg^\ast[1]$. Again, the condition that $Q$ squares to zero is equivalent to the higher Jacobi relations~\eqref{eq:hom_rel} of the original $L_\infty$-algebra. As an example, consider an ordinary Lie algebra $\frg$ generated by degree $0$ generators $t_\alpha$. Then $\wedge^\bullet\frg^\ast[1]$ is spanned by degree $1$ generators $t^\alpha$ and the differential $Q$ acts according to
\begin{equation}\label{eq:ord_diff_ce}
Q t^\alpha = -\tfrac12 f^\alpha_{\beta\gamma} t^\beta t^\gamma~,
\end{equation}
where $f^\alpha_{\beta\gamma}$ are the structure constants of the Lie algebra $\frg$. The condition $\smash{Q^2}=0$ corresponds to $\smash{f^\alpha_{\gamma\left[\beta\right.}f^\gamma_{\left.\phantom{\beta}\!\!\!\!\!\delta\epsilon\right]}}= 0$, which is just the ordinary Jacobi identity.

This readily generalizes to arbitrary $n$-term $L_\infty$-alge\-bras and, altogether, yields a differential graded commutative algebra which is also known as the \textit{Chevalley-Eilenberg algebra} $\ce(\frg)=(\wedge^\bullet\frg^\ast[1],Q)$ of the $L_\infty$-algebra $\frg$. 

Given a Chevalley--Eilenberg algebra $\ce(\frg)$ we can, furthermore, define the concept of a Weil algebra by adding another shifted copy of $\frg$. Explicitly, the \textit{Weil algebra} $\weil(\frg)$ of an $n$-term $L_\infty$-algebra $\frg$ is defined to be
\begin{equation}
\weil(\frg) \coloneqq \big(\wedge^\bullet(\frg^\ast[1]\oplus\frg^\ast[2]), \left.Q_\weil\right|_{\frg^\ast[1]} = Q_{\ce} + \sigma\,\big)~,
\end{equation}
where $\sigma:\frg^\ast[1]\to\frg^\ast[2]$ is the shift isomorphism that identifies elements of $\frg^\ast[1]$ with their shifted copies in $\frg^\ast[2]$. The requirement that $Q_\weil$ squares to zero consequently implies that
\begin{equation}
\left.Q_\weil\right|_{\frg^\ast[2]} = - \sigma Q_\ce \sigma^{-1}~.
\end{equation}
For an ordinary Lie algebra $\frg$ this modifies the above differential to be given by
\begin{equation}\label{eq:ord_lie_alg_diff_weil}
Q t^\alpha = -\tfrac12 f^\alpha_{\beta\gamma} t^\beta t^\gamma + r^\alpha \eand
Q r^\alpha = -f^\alpha_{\beta\gamma} t^\beta r^\gamma~,
\end{equation}
where again $\smash{f^\alpha_{\beta\gamma}}$ are the structure constants of $\frg$ and $r^\alpha$ is the generator in the shifted space $\frg^\ast[2]$. 

The Weil algebra $\weil(\frg)$ straightforwardly projects onto the Chevalley--Eilenberg algebra $\ce(\frg)$,
\begin{equation}
\begin{tikzcd}[arrow style=tikz,>=angle 45]
\ce(\frg) &\weil(\frg)\rlap{~,}\arrow[l,two heads,swap,"p"]
\end{tikzcd}
\end{equation}
which is a sequence that can be extended by the invariant polynomials of an $L_\infty$-algebra $\frg$. Such an \textit{invariant polynomial} is an element $P\in\left.\weil(\frg)\right|_{\wedge^\bullet(\frg^\ast[2])}$, which sits entirely in the shifted copy $\frg^\ast[2]$ inside the Weil algebra and is additionally closed under $Q_\weil$ or, more generally, whose image under $Q_\weil$ also entirely lies in $\wedge^\bullet(\frg^\ast[2])$, cf.~\cite{Sati:2008eg}. Here, we will focus on the closed invariant polynomials and consider only horizontal equivalence classes, where two invariant polynomials $P_1$ and $P_2$ are \textit{horizontally equivalent} if we have $P_1-P_2 = Q\tau$ for an element $\tau\in\ker(p)$. 

For Lie algebras this definition corresponds to the ordinary notion of invariant polynomials and it is these, that, in Chern--Weil theory, are identified with the characteristic classes of the group $\sG$ integrating $\frg$.
The horizontal equivalence classes of the invariant polynomials form the \textit{algebra of invariant polynomials} $\inv(\frg)$ that sits in the sequence
\begin{equation}\label{eq:short_exact_seq_cwi}
\begin{tikzcd}[arrow style=tikz,>=angle 45]
\ce(\frg) &\weil(\frg)\arrow[l,two heads,swap,"p"] & \inv(\frg)\rlap{~,}\arrow[l,swap,hook',"i"]
\end{tikzcd}
\end{equation}
where $i$ is the natural inclusion of $\inv(\frg)$ in $\weil(\frg)$. This sequence will feature prominently in the following.

An immediate advantage of viewing $L_\infty$-algebras as differential graded commutative algebras is that it is now clear what a morphism between arbitrary $n$-term $L_\infty$-algebras should be, i.e., a morphism between the corresponding differential graded commutative algebras. More precisely, a morphism between Chevalley--Eilenberg algebras $\ce(\frg)$ and $\ce(\frg')$ is given by a map $\Phi:\wedge^\bullet(\frg^\ast[1]) \to \wedge^\bullet(\frg^{\prime\ast}[1])$ of degree $0$, that preserves the product structure and respects the differential, i.e.\ $\Phi \circ Q = Q^\prime \circ \Phi$. 

By construction of the Weil algebra, a morphism between Chevalley--Eilenberg algebras also entirely determines a morphism between the corresponding Weil algebras. Given a morphism $\Phi:\ce(\frg)\to\ce(\frg\prime)$, defined on generators $a\in\frg^\ast[1]$ as $a\mapsto \Phi(a)$, we can extend it to a morphism $\hat{\Phi}:\weil(\frg)\to\weil(\frg\prime)$ using  $\sigma a \mapsto \sigma \Phi(a)$. It is straightforward to check that this still respects the differentials and, thus, is a morphism of Weil algebras.

Having defined $L_\infty$-algebras and their morphisms, we now need an appropriate notion of 2-morphisms in order to find categorically equivalent $L_\infty$-algebras. While, in principle, we can identify categorically equivalent $L_\infty$-algebras by finding a morphism between them that induces an isomorphism on the cohomology of the underlying complex~\cite{Kontsevich:1997vb}, we are, for our purposes, going to need an explicit form for the relevant 2-morphisms. An explicit definition has been given in~\cite{Sati:2008eg}, which we will recall here.

Said definition relies on the fact that the Weil algebra $\weil(\frg)$ of an $n$-term $L_\infty$-algebra is naturally isomorphic to the corresponding \textit{free algebra} $\free(\frg)$, which is given by $\free(\frg)\coloneqq (\wedge^\bullet(\frg^\ast[1]\oplus\frg^\ast[2]),Q_F = \sigma)$, where $\sigma:\frg^\ast[1]\to\frg^\ast[2]$ is again the shift isomorphism. To see that this is isomorphic to the Weil algebra consider the morphism
\begin{equation}\label{eq:weil_free_iso}
\begin{aligned}
\Phi: \free(\frg)\to\weil(\frg),\quad &a\mapsto a~,\\
& \sigma a \mapsto Q_\weil a~,\\
\Phi^{-1}: \weil(\frg)\to\free(\frg),\quad & a \mapsto a~,\\
&\sigma a \mapsto \sigma a - Q_\ce a~,
\end{aligned}
\end{equation}
where $a$ denotes the elements in $\frg^\ast[1]$ and $\sigma a$ the corresponding elements in $\frg^\ast[2]$. The compositions $\Phi\circ\Phi^{-1}$ and $\Phi^{-1}\circ\Phi$ yield the identity and it is straightforward to check that the differentials are respected.

Using this isomorphism, we can define 2-morphisms for general Chevalley--Eilenberg algebras: given two morphism $\Phi$ and $\Psi$ between Chevalley--Eilenberg algebras $\ce(\frg)$ and $\ce(\frh)$ we first define a $2$-morphism on the generators of the free algebra $\free(\frg)$ isomorphic to $\weil(\frg)$, the Weil algebra corresponding to $\ce(\frg)$. More explicitly, a \textit{2-morphism} $\eta$ between $\Phi$ and $\Psi$, i.e.
\begin{equation}
\begin{tikzcd}[arrow style=tikz, >=angle 45]
\ce(\frh) &\ce(\frg)\rlap{~,}\arrow[l,bend left=50,"\Psi"{name=U,below}]\arrow[l,bend right=50,"\Phi"{name=D},swap]
\arrow[Rightarrow,"\eta", from=D, to=U,start anchor={[yshift=-1ex]},end anchor={[yshift=1ex]}]
\end{tikzcd}
\end{equation}
is given by a degree $-1$ map
\begin{equation}
\eta: \frg^\ast[1]\oplus\frg^\ast[2]\to \ce(\frh)
\end{equation}
defined on the generators $\frg^\ast[1]\oplus\frg^\ast[2]$ of the free algebra $\free(\frg)$ such that on those generators $\Psi-\Phi=[Q,\eta]$. This morphism is extended to the full space $\free(\frg)$ using the formula
\begin{equation}
\begin{aligned}\label{eq:expansion_formula}
\eta: l_1&\cdots l_n\mapsto\tfrac{1}{n!} \sum\limits_{\sigma \in S_n} \chi(\sigma) \sum\limits_{k=1}^n (-1)^{\sum\limits_{i=1}^{k-1}\left|l_{\sigma(i)}\right|} \\
&\psi(l_{\sigma(1)}\cdots l_{\sigma(k-1)})\eta(l_{\sigma(k)})\phi(l_{\sigma(k+1)}\cdots l_{\sigma(n)})~,
\end{aligned}
\end{equation}
where $l_i \in \frg^\ast[1]\oplus\frg^\ast[2]$. This guarantees that $\psi-\phi=[Q,\eta]$ on the whole of $\free(\frg)$. Additionally, we require that $\eta$, when viewed as a morphism out of $\weil(\frg)$, vanishes on the generators in the shifted copy inside $\weil(\frg)$. More explicitly, when comparing with the isomorphism~\eqref{eq:weil_free_iso}, this means that $\eta: \free(\frg)\to \ce(\frh)$ defines a map on the generators $a$ and $Q_\weil a$ inside $\weil(\frg)$ in such a way that $\eta$ vanishes on all $\sigma a \in \weil(\frg)$. In summary, we have a diagram
\begin{equation}
\begin{tikzcd}[arrow style=tikz,>=angle 45]
&\ce(\frg)\arrow[dl,bend right=30,"\phi",swap]&&\frg^\ast[2]\\
\ce(\frh) &&\weil(\frg)\arrow[ur,hookleftarrow]\arrow[ul,"i^\ast"{name=U,below}]\arrow[dl,"i^\ast"]&\free(\frg)\rlap{~,}\arrow[l,swap,"\cong"]\\
&\ce(\frg)\arrow[ul,bend left=30,"\psi"{name=D}]&& \arrow[Rightarrow,"\eta",from=U,to=D,start anchor={[xshift=-1ex]},end anchor={[yshift=1ex,xshift=1ex]}]
\end{tikzcd}
\end{equation}
where $\eta$ vanishes along $\frg^\ast[2]\hookrightarrow\weil(\frg)$. Note, that this definition extends to morphisms between Weil algebras as $\weil(\frg)$ can be seen as the Chevalley--Eilenberg algebra $\ce(\frg\overset{\text{id}}{\to}\frg[-1])$.

It is instructive to spell out what this definition means in the example of $2$-term $L_\infty$-algebras $\frg$ and $\frh$. Let $t^\alpha\,(t^{\prime\alpha})$ and $b^a\,(b^{\prime a})$ be the generators of $\frg\,(\frh)$ in degrees~1~and~2, respectively. A generic degree $0$ map $\Phi$ is then given by
\begin{equation}\label{eq:gendegzeromap}
\begin{aligned}
\Phi (t^\alpha) &= \Phi^\alpha_\beta t^{\prime\beta}~,\\
\Phi (b^a) &= \Phi^a_b b^{\prime b} + \tfrac12 \Phi^a_{\alpha\beta} t^{\prime\alpha}t^{\prime\beta}~,
\end{aligned}
\end{equation}
while a generic degree $-1$ map $\eta$ can be written as
\begin{equation}
\begin{aligned}
\eta (t^\alpha) &= 0~,\\
\eta (b^a) &= \eta^a_\alpha t^\alpha~.
\end{aligned}
\end{equation}
The requirement that $\eta$ vanishes along $\frg^\ast[2]\subset\weil(\frg)$ together with the formula~\eqref{eq:expansion_formula} also defines $\eta$ on $Qt^\alpha$ and $Qb^a$, which we use to calculate
\begin{equation}\label{eq:qetaresult}
\begin{aligned}
[Q,\eta] t^\alpha &= - f^\alpha_a\eta^a_\beta \, t^{\prime\beta}~,\\
[Q,\eta] b^a &= -\tfrac12 \eta^a_\alpha f^{\prime\alpha}_{\beta\gamma} \,t^{\prime\beta}\wedge t^{\prime\gamma} -\eta^a_\alpha f^{\prime\alpha}_b \,b^{\prime b} \\
&\phantom{{}={}}+ \tfrac12 f^a_{\alpha b} \eta^b_\gamma(\Psi^\alpha_\beta + \Phi^\alpha_\beta) \,t^{\prime\beta}\wedge t^{\prime\gamma}~,
\end{aligned}
\end{equation}
where $f\,(f')$ are the generic structure constants of the 2-term $L_\infty$-algebras. The requirement that $\psi-\phi=[Q,\eta]$ can then be read of from~\eqref{eq:gendegzeromap} and~\eqref{eq:qetaresult} to translate to
\begin{equation}
\begin{aligned}
\Phi^\alpha_\beta-\Psi^\alpha_\beta &= -f^\alpha_a \eta^a_\beta~,\\
\Phi^a_b - \Psi^a_b &= - \eta^a_\alpha f^{\prime\alpha}_b~,\\
\Phi^a_{\left[\beta\gamma\right]}-\Psi^a_{\left[\beta\gamma\right]} &= -\eta^a_\alpha f^{\prime\alpha}_{\left[\beta\gamma\right]}+f^a_{\alpha b} (\Psi + \Phi)^\alpha_{\left[\beta\right.}\eta^b_{\left.\gamma\right]}~.
\end{aligned}
\end{equation}
This is precisely the more familiar condition for 2-mor\-phisms as given in~\cite{Baez:2003aa}, also cf.~\cite[Appendix A]{Sati:2008eg}.

\section{Models for the string Lie 2-algebra}\label{sec:models_string_algebra}

The case of interest to this discussion is a 2-term $L_\infty$-algebra known as the string Lie 2-algebra\footnote{We will refer to the string Lie 2-algebra in all its incarnations simply as the string algebra.}. For a discussion of reasons why this algebra is of interest we refer to~\cite{Saemann:2017rjm,Saemann:2017zpd}, also cf.~\cite{contrib:saemann}. There are different models for the string algebra and here we will focus on two extremes: a skeletal model, that is, a 2-term $L_\infty$-algebra with vanishing $\mu_1$, and a strict model, that is, a 2-term $L_\infty$-algebra with vanishing $\mu_3$.

The \textit{skeletal model of the string algebra}, denoted as $\astringsk(\frg)$, is a 2-term $L_\infty$-algebra given by
\begin{equation}
\astringsk(\frg) = \big(\,\,\frg \overset{0}{\longleftarrow} \FR\,\,\big)~,
\end{equation}
together with non-trivial brackets
\begin{equation}\label{eq:skel_string_algebra_brackets}
\begin{aligned}
 \mu_2&:\frg\wedge \frg\rightarrow \frg~,~~~&\mu_2(l_1,l_2)&=[l_1,l_2]~,\\
 \mu_3&:\frg\wedge \frg\wedge \frg\rightarrow \FR~,~~~&\mu_3(l_1,l_2,l_3)&=(l_1,[l_2,l_3])~,
\end{aligned}
\end{equation}
where $[-,-]$ is the commutator and $(-,-)$ is the Killing form of $\frg$. The corresponding Weil algebra is given by generators $t^\alpha, b^a$ of degree~1~and~2, respectively, together with their shifted copies $r^\alpha = \sigma t^\alpha$ and $h^a = \sigma b^a$ of degree~2~and~3. The differential belonging to~\eqref{eq:skel_string_algebra_brackets} is then given by
\begin{equation}\label{eq:skel_string_algebra_differential}
\begin{aligned}
Qt^\alpha &= -\tfrac12 f^{\alpha}_{\beta\gamma} t^\beta t^\gamma + r^\alpha~,\\
Qb^a &= -\tfrac16 f^a_{\alpha\beta\gamma} t^\alpha t^\beta  t^\gamma + h^a~,\\
Qr^\alpha &= -f^\alpha_{\beta\gamma} t^\beta  r^\gamma~,\\
Qh^a &= \tfrac12 f^a_{\alpha\beta\gamma} t^\alpha t^\beta  r^\gamma~,
\end{aligned}
\end{equation}
where $f^\alpha_{\beta\gamma}$ and $f^a_{\alpha\beta\gamma}$ are the structure constants corresponding to $\mu_2$ and $\mu_3$, respectively.

The strict model, on the other hand, is referred to as the \textit{loop model of the string algebra}, denoted as $\astringl(\frg)$, which is a 2-term $L_\infty$-algebra based on the space
\begin{equation}
\astringl(\frg) = \big(\,\,P_0\frg\overset{\mu_1}{\longleftarrow}\Omega\frg\oplus\FR\,\,\big)~,
\end{equation}
where $P_0\frg$ and $\Omega\frg$ are the spaces of based paths and loops in $\frg$, respectively. The non-trivial brackets are
\begin{align}\label{eq:loop_string_algebra_brackets}
\begin{aligned}
\mu_1&:\Omega\frg\oplus\FR \rightarrow P_0\frg~,&\mu_1((\lambda,r)) &= \lambda~,\\
\mu_2&:P_0\frg\wedge P_0\frg\rightarrow P_0\frg~,&\mu_2(\gamma_1,\gamma_2)&=\![\gamma_1\!,\!\gamma_2]~,\\
\mu_2&:P_0\frg\otimes(\Omega\frg\oplus\FR)\rightarrow \Omega\frg\oplus\FR~,&&
\end{aligned}\\\negthickspace\negmedspace
\mu_2\big(\gamma,(\lambda,r)\big)=\left([\gamma,\lambda]\; ,\; -2\int_0^1 \dd\tau \left(\gamma(\tau),\dder{\tau}\lambda(\tau)\right)\right)~,\nonumber
\end{align}
where $[-,-]$ is the pointwise commutator and $(-,-)$ the pointwise Killing form of $P_0\frg$. For the corresponding Weil algebra we introduce pointwise generators of the dual given by $t^{\alpha\tau}$ in degree $1$ and $(b^{a\tau},b^a)$ in degree $2$, together with their shifted copies $r^{\alpha\tau}$ and $(h^{a\tau},h^a)$ of degrees~2~and~3~, respectively. As this is an infinite-dimensional case, we need to additionally introduce generators $\dot{b}^{a\tau}$ and $\dot{h}^{a\tau}$, which are the duals corresponding to derivatives of paths. This will prove sufficient for all subsequent calculations. The differential of the Weil algebra can then be written as
\begin{equation}
\begin{aligned}
Q t^{\alpha\tau} &= -\tfrac12 f^{\alpha}_{\beta\gamma,\tau} t^{\beta\tau} t^{\gamma\tau} -f^\alpha_{a,\tau} b^{a\tau}+ r^{\alpha\tau}~,\\
Q b^{a\tau} &= -f^a_{\beta b,\tau} t^{\beta\tau} b^{b\tau} + h^{a\tau}~,\\
Q b^{a\phantom{\tau}} &= - f^{\prime a}_{\beta b,\tau} t^{\beta\tau}\dot{b}^{b\tau} + h^a~,\\
Q r^{\alpha\tau} &= -f^\alpha_{\beta\gamma,\tau} t^{\beta\tau} r^{\gamma\tau} + f^\alpha_{a,\tau} h^{a\tau} ~,\\
Q h^{a\tau} &= f^{a}_{\alpha b,\tau}r^{\beta\tau} b^{b\tau} - f^a_{\beta b,\tau} t^{\beta\tau} h^{b\tau}~,\\
Q h^{a\phantom{\tau}} &= f^{\prime a}_{\alpha b,\tau}r^{\beta\tau} \dot{b}^{b\tau} - f^{\prime a}_{\beta b,\tau} t^{\beta\tau} \dot{h}^{b\tau}~,
\end{aligned}
\end{equation}
where $f^\alpha_{a,\tau}$ corresponds to $\mu_1$, $\smash{f^\alpha_{\beta\gamma,\tau}}$ corresponds to the pointwise commutator and $\smash{(f^a_{\alpha b,\tau}, f^{\prime a}_{\alpha b,\tau})}$ corresponds to the different parts of the mixed $\mu_2$ in~\eqref{eq:loop_string_algebra_brackets}.

In~\cite{Baez:2005sn} it was shown that these two different models for the string algebra are categorically equivalent and, thus, should be interchangeable.

\section{Higher gauge theory from morphisms}

Viewing $L_\infty$-algebras in terms of differential graded commutative algebras provides another advantage: as the de Rham complex of differential forms also form such an algebra it offers a unifying framework for the ingredients of higher gauge theory and, thus, enables an elegant description of the local data of connections together with corresponding curvatures, Bianchi identities and gauge transformations. This approach is a generalization of ideas by Cartan~\cite{Cartan:1949aaa,MR0042426} and Atiyah~\cite{Atiyah:1957}, partially due to~\cite{Bojowald:0406445,Kotov:2010wr,Gruetzmann:2014ica} and, to its full extent, due to~\cite{Sati:2008eg}.

In this framework, a flat connection on an open set $U\subset\FR^n$ is given locally by a morphism from the Chevalley--Eilenberg algebra of the relevant $L_\infty$-algebra $\frg$ to the de Rham complex $(\Omega^\bullet(U),\dd=\dd x^\mu \partial_\mu)$, i.e.
\begin{equation}
\begin{tikzcd}[arrow style=tikz,>=angle 45]
\Omega^\bullet(U) & \text{CE}(\frg)\rlap{~.}\arrow[l,"A",swap]
\end{tikzcd}
\end{equation}
For an ordinary Lie algebra with differential as given in~\eqref{eq:ord_diff_ce} the morphism $A$ acts as $t^\alpha\mapsto A^\alpha_\mu \dd x^\mu$. Requiring that $A$ should respect the differentials leads to the equation
\begin{equation}
\dd A^\alpha = -\tfrac12 f^\alpha_{\beta\gamma} A^\beta \wedge A^\gamma~,
\end{equation}
which corresponds to $A$ being flat, i.e.\ to its curvature vanishing.

To allow for more general connections we can replace the Chevalley--Eilenberg algebra $\ce(\frg)$ with its corresponding Weil algebra $\weil(\frg)$. As discussed above, by the construction of the Weil algebra, the morphism $A$ induces a morphism $\CF$ on the shifted copy of $g$ so that a connection is now encoded
as
\begin{equation}
\begin{tikzcd}[arrow style=tikz,>=angle 45]
\Omega^\bullet(U) & \text{W}(\frg)\rlap{~.}\arrow[l,"{(A, \CF)}",swap]
\end{tikzcd}
\end{equation}
In the case of an ordinary Lie algebra's Weil algebra $\weil(\frg)$ with a differential as in~\eqref{eq:ord_lie_alg_diff_weil} the morphisms act as $t^\alpha \mapsto A^\alpha_\mu \dd x^\mu$ and $r^\alpha \mapsto \tfrac12\CF^\alpha_{\mu\nu} \dd x^\mu \wedge \dd x^\nu$ and the condition that the differential is respected consequently translates to
\begin{equation}
\begin{aligned}
\CF^\alpha &= \dd A^\alpha + \tfrac12 f^\alpha_{\beta\gamma} A^\beta \wedge A^\gamma~,\\
\dd \CF^\alpha &= -f^\alpha_{\beta\gamma} A^\beta \wedge \CF^\gamma~.
\end{aligned}
\end{equation}
Thus, we not only incorporate a non-vanishing curvature $\CF$ but also conveniently encode its Bianchi identity. This procedure readily generalizes to arbitrary $n$-term $L_\infty$-algebras.

Furthermore, gauge transformations can be encoded in flat homotopies between two such gauge configurations~\cite{Fiorenza:2010mh}, i.e.\ in morphisms
\begin{equation}
\begin{tikzcd}[arrow style=tikz,>=angle 45]
\Omega^\bullet(U\times I) & \text{W}(\frg)\rlap{~,} \arrow[l,"{(A,\CF)}",swap]
\end{tikzcd}
\end{equation}
where $I=[0,1]$ denotes the interval and $\CF$ vanishes on those additional directions. Denoting the coordinate in the additional direction by $\rho$, the differential on $\Omega^\bullet(U\times I)$ is given by $\dd_{\text{ext}} = \dd x^\mu \partial_\mu + \dd \rho \der{\rho}$. Then, for an ordinary Lie algebra $\frg$, such a morphism is defined on coordinates as $t^\alpha\mapsto A^\alpha_\mu \dd x^\mu + \lambda^\alpha \dd \rho$ and $r^\alpha\mapsto \tfrac12\CF^\alpha_{\mu\nu}\dd x^\nu\wedge\dd x^\nu$ and respecting the differentials translates to
\begin{equation}\label{eq:generic_gauge_trafo}
\begin{aligned}
\CF^\alpha &= \dd A^\alpha + \tfrac12 f^\alpha_{\beta\gamma} A^\beta \wedge A^\gamma~,\\
\dd \CF^\alpha &= -f^\alpha_{\beta\gamma} A^\beta \wedge \CF^\gamma~,\\
\delta_\lambda A^\alpha &= \dd \lambda^\alpha + \tfrac12 f^\alpha_{\beta\gamma} A^\beta \lambda^\gamma~,\\
\delta_\lambda \CF^\alpha &= f^\alpha_{\beta\gamma} \CF^\beta \lambda^\gamma~,
\end{aligned}
\end{equation}
where the first two lines are as before and the additional equations are the familiar expressions for an infinitesimal gauge transformation with gauge parameter $\lambda$. Again, this procedure readily generalizes to higher, arbitrary $n$-term $L_\infty$-algebras.

Following the ideas in~\cite{Sati:2008eg}, we can also look at global properties and consider what the usual Ehresmann conditions of a connection imply for the morphisms above. Given a principal bundle $\pi:P\twoheadrightarrow M$, let $\Omega^\bullet_{\text{vert}}(P)$ be the algebra of vertical differential forms, that is, the forms that have legs only along the vertical vector fields. More precisely, $\Omega^\bullet_{\text{vert}}(P)$ is given by $\Omega^\bullet(P)$ modulo the ideal generated by the forms that vanish when restricted to the kernel of $\pi_\ast:\Gamma(TP)\to\Gamma(TM)$. First, consider the square
\begin{equation}\label{eq:first_ehresmann_square}
\begin{tikzcd}[arrow style=tikz,>=angle 45]
\Omega_\text{vert}^\bullet(P) & \ce(\frg) \arrow[l,"A_\text{vert}",swap] \\
\Omega^\bullet(P) \arrow[u] & \weil(\frg)\rlap{~.} \arrow[u,"p",two heads,swap]\arrow[l,"{(A,\CF)}",swap]
\end{tikzcd}
\end{equation}
The commutativity of this square implies that the composite morphism $\text{W}(\frg)\to\Omega^\bullet(P)\to\Omega_\text{vert}^\bullet(P)$, i.e.\ the connection $A$ along the fibres of $P$, factors along $\text{CE}(\frg)$ to give a map $A_\text{vert}$. Being a morphism out of $\text{CE}(\frg)$, this therefore means that $A_\text{vert}$ has vanishing curvature. This is, as usual, implied by the first Ehresmann condition as it requires the connection to simply be the Maurer--Cartan form along the fibres, which has a vanishing curvature. As such, the first Ehresmann condition implies the commutativity of the above square.

Second, recall that the second Ehresmann condition implies that the characteristic classes $<\CF>$ of the principal bundle descend down to global forms on the base manifolds. In terms of the above morphisms, this leads to the commutativity of the square
\begin{equation}\label{eq:second_ehresmann_square}
\begin{tikzcd}[arrow style=tikz, >=angle 45]
\Omega^\bullet (P) & \weil(\frg) \arrow[l,"{(A,\CF)}",swap] \\
\Omega^\bullet (M) \arrow[u] & \inv(\frg)\rlap{~,}\arrow[l,"<\CF>",swap]\arrow[u,"i",swap,hook']
\end{tikzcd}
\end{equation}
as it is the invariant polynomials that are identified with the characteristic classes of the bundle.

Combining~\eqref{eq:first_ehresmann_square} and~\eqref{eq:second_ehresmann_square}, we have that for a connection $A$ the diagram
\begin{equation}\label{eq:g_connection}
\begin{tikzcd}[arrow style=tikz,>=angle 45]
\Omega^\bullet_{\rm{vert}}(P) & \ce(\frg) \arrow[l,"A_{\rm vert}",swap]\\
\Omega^\bullet(P) \arrow[u] & \weil(\frg) \arrow[l,"{(A,\CF)}",swap]\arrow[u,"p",swap, two heads] \\
\Omega^\bullet(M)\arrow[u] & \inv(\frg)\arrow[u,"i",swap,hook']\arrow[l,"<\CF>",swap]
\end{tikzcd}
\end{equation}
commutes. This condition will prove crucial in calculating the twists in the following.

\section{The fake flatness condition}

The above formalism yields naive expressions for the curvatures of the connections for arbitrary $n$-term $L_\infty$-algebras. For a 2-term $L_\infty$-algebra, e.g., one finds a two-form curvature $\CF$ together with a three-form curvature $\CH$ corresponding to the one-form connection $A$ and a two-form connection $B$. In higher gauge theory, one frequently encounters the \textit{fake flatness condition} $\CF=0$, which, e.g., is responsible for rendering higher parallel transport of strings invariant under surface reparameterizations~\cite{Baez:2004in}, see also~\cite{Baez:2010ya,Saemann:2011nb,Saemann:2012uq}. 

However, this condition makes it difficult to find interesting examples of higher gauge theory with relevance to string and M-theory. To see this, consider the example of the models of the string algebra outlined in Section~\ref{sec:models_string_algebra}. In the skeletal model we have potentials $A\in\Omega^1(U,\frg)$ and $B\in\Omega^2(U,\FR)$ and are led to the curvatures
\begin{equation}
\begin{aligned}
\CF &= \dd A + \tfrac12 \mu_2(A,A)~,\\
\CH &= \dd B + \tfrac16 \mu_3(A,A,A)~,
\end{aligned}
\end{equation}
where $\CF\in\Omega^2(U,\frg)$ and $\CH\in\Omega^3(U,\FR)$. Similarly, in the loop model we have potentials $A\in\Omega^1(U,P_0\frg)$ and $B\in\Omega^2(U,\Omega\frg\oplus\FR)$ with curvatures $\CF\in\Omega^2(U,P_0\frg)$ and $\CH\in\Omega^3(U,\Omega\frg\oplus\FR)$ given by
\begin{equation}
\begin{aligned}
\CF &= \dd A + \tfrac12 \mu_2(A,A) + \mu_1(B)~,\\
\CH &= \dd B + \mu_2(A,B)~.
\end{aligned}
\end{equation}

In~\cite{Saemann:2017rjm}, this gauge structure was used to attempt to find a non-abelian analogue of the abelian self-dual string, which is the M-theory analogue of the 't Hooft--Polyakov monopole. In the skeletal case, the condition $\CF = 0$ implies that the connection $A$ is pure gauge and can be gauged away. This, however, reduces the three-form curvature to $\CH=\dd B$, which is just the abelian version. Thus, the fake flatness condition prohibits the construction of a non-abelian analogue. Similar arguments in the loop case lead to the same conclusion. For further details we refer to~\cite{Saemann:2017rjm} or~\cite{contrib:saemann}.

On the other hand, it can be shown that with the above curvatures the fake flatness condition is needed to render the equations gauge covariant and for the descriptions in terms of the two different models to be equivalent, see~\cite{Saemann:2017rjm}. This suggests that the naive curvatures above need to be modified and the algebraic structures are not quite the ones needed for interesting constructions. A way to escape this dilemma is to use twisted Weil algebras of the string algebra, which already appear in~\cite{Sati:2008eg,Fiorenza:2010mh,Sati:2009ic} and have been used in~\cite{Saemann:2017rjm} to successfully construct a non-abelian self-dual string. In the following we will discuss a way to arrive at these twists.

\section{Twisted Weil algebras for the string Lie 2-algebra}

Before discussing the twisted Weil algebras of the string algebra themselves let us introduce a few more terms to simplify the discussion. An element $\mu\in\ce(\frg)$ that closes under $Q_\ce$ is called an \textit{$L_\infty$-algebra cocycle}. Given such a cocycle $\mu$ and an invariant polynomial $P\in\inv(\frg)$ we call an element ${\rm{cs}}\in\weil(\frg)$ that satisfies
\begin{equation}
Q_\weil {\rm cs} \,=\, i (P) \eand
p ({\rm cs}) \,=\, \mu~,
\end{equation}
a \textit{$\frg$-transgression element} or \textit{Chern--Simons element} for $\mu$ and $P$.
If such a transgression element exists for a given cocycle $\mu$ and invariant polynomial $P$, we say, that $\mu$ \textit{transgresses to} $P$ and $P$ \textit{suspends to} $\mu$. Note that given this, $\mu$ is indeed a cocycle:
\begin{equation}
Q_\ce \mu \,=\, Q_\ce p( {\rm cs}) \,=\, p (Q_\weil {\rm cs}) \,=\, p (i(P)) \,=\, 0~,
\end{equation}
where we use the fact that $p$ is a morphism of differential graded algebras and $\im(i)\subset\ker(p)$. The cohomology class of the cocycle $\mu$ is independent of the transgression element ${\rm cs}$ chosen. Indeed, considering $\mu^\prime = \mu + Q_\ce a$ for some $a \in \ce(\frg)$ we have $\mu^\prime = p({\rm cs} + Q_\weil a)$ and $Q_\weil({\rm cs} + Q_\weil a) = Q_\weil {\rm cs} =i(P)$, so that $\mu^\prime$ transgresses to the same invariant polynomial. Therefore, an invariant polynomial that suspends to a coboundary $\mu = Q_\ce a$ also suspends to~0.

\pagebreak
One philosophy in defining higher connections and curvatures is that they should be in a sense a lift of an ordinary connection. It is following this philosophy that will lead us to the twisted Weil algebras of the string algebra. Let us start by focusing on the skeletal model: there is an obvious projection of $\astringsk(\frg)$ down to $\frg$, so that one could desire for a $\astringsk(\frg)$-connection to be a lift of a corresponding $\frg$-connection. That is, one can ask whether or not the diagram~\eqref{eq:g_connection} lifts to a $\astringsk(\frg)$-connection, i.e.\ to
\begin{equation}
\begin{tikzcd}[arrow style=tikz,>=angle 45,column sep=2ex]
\ce(\astringsk(\frg))\arrow[dr,dashed] && \ce(\frg)\arrow[ll,hook']\arrow[dl, "A_{\rm vert}" description] \\
& \Omega^\bullet_{\rm vert}(P) & \\
\weil(\astringsk(\frg))\arrow[uu]\arrow[dr,dashed] && \weil(\frg)\arrow[uu]\arrow[ll,hook']\arrow[dl,"{(A,\CF)}" description] \\
& \Omega^\bullet(P)\arrow[uu,crossing over] & \\
\inv(\astringsk(\frg))\arrow[dr,dashed]\arrow[uu] && \inv(\frg)\arrow[uu]\arrow[ll,hook']\arrow[dl,"<\CF>" description ]\rlap{~.} \\
& \Omega^\bullet(M)\arrow[uu,crossing over] &
\end{tikzcd}
\end{equation}
In general, this is not possible. However, we can instead consider the extended algebra
\begin{equation}
\astringskext(\frg)\, \coloneqq\,\big(\, \astringsk(\frg) \overset{\mu_1}{\longleftarrow} \FR[2]\,\big)~,
\end{equation}
where $\mu_1=\text{id}$ is the only additional structure map. This extended algebra, as can be quickly seen from cohomology, is categorically equivalent to $\frg$ and thus comes with an equivalence $\Phi:\smash{\ce(\astringskext(\frg)) \overset{\sim}{\longrightarrow} \ce(\frg)}$, which we can employ to extend our diagram to
\pagebreak
\begin{equation}
\begin{tikzcd}[arrow style=tikz,>=angle 45,column sep=1.2ex]
&&& \ce(\astringskext(\frg))\arrow[dl,"\sim",swap,outer sep=-2pt]\arrow[dlll] \\
\ce(\astringsk(\frg))\arrow[dr,dashed] && \ce(\frg)\arrow[ll,hook']\arrow[dl, "A_{\rm vert}" description] &\\
& \Omega^\bullet_{\rm vert}(P) &&  \weil(\astringskext(\frg)) \arrow[dl,"\sim",swap,outer sep=-2pt]\arrow[dlll]\arrow[uu,dotted]\\
\weil(\astringsk(\frg))\arrow[uu]\arrow[dr,dashed] && \weil(\frg)\arrow[uu,crossing over]\arrow[ll,hook']\arrow[dl,"{(A,\CF)}" description] &\\
& \Omega^\bullet(P)\arrow[uu,crossing over] &&  \inv(\astringskext(\frg))\arrow[dl,"=",swap,outer sep=-2pt]\arrow[dlll]\arrow[uu,dotted]\\
\inv(\astringsk(\frg))\arrow[dr,dashed]\arrow[uu] && \inv(\frg)\arrow[uu,crossing over]\arrow[ll,hook']\arrow[dl,"<\CF>" description ]\rlap{~.} &\\
& \Omega^\bullet(M)\arrow[uu,crossing over] &&
\end{tikzcd}
\end{equation}
That is, while there may not be a lift to a $\astringsk(\frg)$-connection, we do, at least in principle, always get a $\astringskext(\frg)$-connection. Furthermore, as $\ce(\astringskext(\frg))$ projects down to $\ce(\astringsk(\frg))$, this also conveniently measures the failure to lift this to a $\astringsk(\frg)$-connection object. That is, the obstruction is measured by the non-triviality of the component of $A\circ\Phi$ on the extra generator in the additional $\FR$, which needs to vanish in order for the lift to exist.

However, there is a subtlety here, that is crucial: the partial diagram
\begin{equation}\label{eq:non_comm_diagram}
\begin{tikzcd}[arrow style=tikz,>=angle 45]
\ce(\frg) & \ce(\astringskext(\frg))\arrow[l] \\
\weil(\frg)\arrow[u] & \weil(\astringskext(\frg))\arrow[l]\arrow[u,dotted] \\
\inv(\frg)\arrow[u] & \inv(\astringskext(\frg)) \arrow[l]\arrow[u,dotted]
\end{tikzcd}
\end{equation}
does not commute and the $\astringskext(\frg)$-connection is therefore not yet well-defined. 

In order to see this, let us discuss the involved concepts and morphisms in more detail. Let the additional coordinates of degree~3~and~4~ in $\weil(\astringsk(\frg))$ be denoted by $c^\mu$ and $g^\mu$, respectively, and let the differential in~\eqref{eq:skel_string_algebra_differential} be modified to
\begin{equation}
\begin{aligned}
Q t^\alpha &= -\tfrac12 f^\alpha_{\beta\gamma} t^\beta t^\gamma + r^\alpha~,~~Q r^\alpha = -f^\alpha_{\beta\gamma} t^\alpha r^\beta~,\\
Q b^a &= -\tfrac16 f^a_{\alpha\beta\gamma} t^\alpha t^\beta t^\gamma -f^a_\mu c^\mu+ h^a~,~~Q c^\mu = g^\mu\\
Q h^a &= \tfrac12 f^a_{\alpha\beta\gamma} t^\alpha t^\beta r^\gamma + f^a_\mu g^\mu~,~~ Qg^\mu=0~,
\end{aligned}
\end{equation}
where $f^a_\mu$ is the identity.

Consider first the invariant polynomials of $\astringsk(\frg)$ itself --- they agree with the ones of $\frg$ with one exception. We have that $\mu^a = -\tfrac16 f^a_{\alpha\beta\gamma} t^\alpha t^\beta t^\gamma$ is a cocycle, as $Q_\ce \mu ^a = 0$. In defining $\astringsk(\frg)$ we introduced the additional generator $b^a$ that explicitly turns $\mu^a$ into a coboundary, i.e.\ $Qb^a = \mu^a$. Therefore, the invariant polynomial $P^a$ that $\mu^a$ transgresses to now suspends to and, thus, is horizontally equivalent to~0. A transgression element for $P^a$ and $\mu^a$ is given by
\begin{equation}
{\rm cs^a} = -\tfrac16 f^a_{\alpha\beta\gamma} t^\alpha t^\beta t^\gamma + \kappa^a_{\alpha\beta} t^\alpha r^\beta~,
\end{equation}
where $\kappa^a_{\alpha\beta}$ encodes the Killing form, so that $f^a_{\alpha\beta\gamma} = \kappa^a_{\alpha\delta}f^\delta_{\beta\gamma}$. A quick calculation shows that this leads to the invariant polynomial
\begin{equation}
P^a = Q_\weil ({\rm cs}^a) = \kappa^a_{\alpha\beta} r^\alpha r^\beta~.
\end{equation}
Thus, $\inv(\astringsk)$ consists of the invariant polynomials in $\inv(\frg)$ barring $P^a= \kappa^a_{\alpha\beta} r^\alpha r^\beta$. 

For $\inv(\astringskext(\frg))$ this situation changes as we introduce an additional generator $c^\mu$ which comes with the additional invariant polynomial $g^\mu$, as $Q g^\mu =0$. We then have
\begin{equation}
P^a = \kappa^a_{\alpha\beta} r^\alpha r^\beta = Q( -h^a + \kappa^a_{\alpha\beta} t^\alpha r^\beta) + f^a_\mu g^\mu~, 
\end{equation}
so that, as $ -h^a + \kappa^a_{\alpha\beta} t^\alpha\wedge r^\beta$ is in $\ker(p)$, now $P^a$ no longer is horizontally equivalent to~0~but rather to the new invariant polynomial $g^\mu$. As such, this restores the missing invariant polynomial and $\inv(\astringskext(\frg))$ is isomorphic to $\inv(\frg)$.

Keeping this in mind, we turn to the explicit form of the equivalences between $\frg$ and $\astringskext(\frg)$. The relevant morphisms are given by
\begin{equation}
\begin{aligned}
\Phi:\weil&(\astringskext(\frg))\to\weil(\frg)~, & \!\!\!\!\!\!\Psi: \weil&(\frg)\to\weil(\astringskext(\frg))~,\\
t^\alpha &\mapsto t^{\prime\alpha}~,~~r^\alpha\mapsto r^{\prime\alpha}~,&t^{\prime\alpha} &\mapsto t^\alpha~,~~r^{\prime\alpha}\mapsto r^\alpha~, \\
b^a &\mapsto 0~,~~ h^a\mapsto 0~, &&\\
c^\mu &\mapsto f^\mu_a \mu^{\prime a}~,&&\\
g^\mu &\mapsto \sigma (f^\mu_a \mu^{\prime a})~.&&
\end{aligned}
\end{equation}
where we label the coordinates in $\weil(\frg)$ as $t^{\prime\alpha}$ and $r^{\prime\alpha}$. It can be checked that $\Phi$ and $\Psi$ respect the differentials and, hence, are indeed morphisms of Weil algebras. Furthermore, $\Phi\circ\Psi$ yields the identity and $\Psi\circ\Phi$ can be connected to the identity via the 2-morphism
\begin{equation}\label{eq:two_morphism}
\begin{aligned}
\eta:\weil&(\astringskext(\frg))\longrightarrow\weil(\astringskext(\frg))~,\\
t^\alpha &\mapsto 0~,~~b^a \mapsto 0~,~~c^\mu \mapsto -f^\mu_a b^a~.\\
r^\alpha &\mapsto 0~,~~h^a \mapsto 0~,~~g^\mu \mapsto f^\mu_a h^a~.
\end{aligned}
\end{equation}
This is indeed a 2-morphism as outlined above and, thus, $\weil(\astringskext(\frg))$ is equivalent to $\weil(\frg)$.

We are now in a position to see how the diagram~\eqref{eq:non_comm_diagram} fails to commute: on the one hand, the invariant polynomial $g^\mu$ in $\inv(\astringskext(\frg))$ is identified with $\kappa_{\alpha\beta} r^{\prime\alpha}r^{\prime\beta}$ in $\inv(\frg)$ and then mapped to the corresponding element in $\weil(\frg)$. On the other hand, $\Phi$ maps $g^\mu\in\weil(\astringskext(\frg))$ to $\sigma(f^\mu_a \mu^{\prime a}) = -\tfrac12  f^\mu_a f^a_{\alpha\beta\gamma} t^{\prime\alpha} t^{\prime\beta} r^{\prime\gamma}$ in $\weil(\frg)$, which clearly is not an invariant polynomial. 

To alleviate this problem we can modify the Weil algebra differential to be given by
\begin{equation}
\begin{aligned}
Q t^\alpha &= -\tfrac12 f^\alpha_{\beta\gamma} t^\beta t^\gamma + r^\alpha~,~~Q r^\alpha = -f^\alpha_{\beta\gamma} t^\alpha r^\beta~,\\
Q b^a &= -\tfrac16 f^a_{\alpha\beta\gamma} t^\alpha t^\beta t^\gamma + \kappa^a_{\alpha\beta} t^\alpha r^\beta -f^a_\mu c^\mu+ h^a~,\\
Q h^a &= -\kappa^a_{\alpha\beta} r^\alpha r^\beta+ f^a_\mu g^\mu~,~~Q c^\mu = g^\mu~,~~ Qg^\mu=0~,
\end{aligned}
\end{equation}
so that $Q b^a$ is modified to contain the Chern--Simons element ${\rm cs}^a$ and, in turn, $Q h^a$ is given by the invariant polynomial $P^a$. This is the \textit{twisted Weil algebra in the skeletal model} and we denote the algebra with the modified Weil algebra by $\twastringsk(\frg)$. Its Chevalley--Eilenberg algebra remains unaffected as the modification $\kappa^a_{\alpha\beta} t^\alpha\wedge r^\beta$ lives in the kernel $\ker(p)$ and we have 
\begin{equation}
\ce(\twastringsk(\frg)) = \ce(\astringskext(\frg))~.
\end{equation}
Furthermore, the invariant polynomials in $\inv(\twastringsk(\frg))$ remain unaffected and, again, are isomorphic to those in $\inv(\frg)$. Additionally, the Weil algebra $\weil(\twastringsk(\frg))$ is still equivalent to both $\weil(\frg)$ and $\weil(\astringskext(\frg))$. Explicitly, the equivalences $\Phi$ and $\Psi$ are modified to be
\begin{equation}\label{eq:final_equivalence}
\begin{aligned}
\Phi:\weil&(\twastringsk(\frg))\to\weil(\frg)~, & \!\!\!\!\!\!\Psi: \weil&(\frg)\to\weil(\twastringsk(\frg))~,\\
t^\alpha &\mapsto t^{\prime\alpha}~,~~r^\alpha\mapsto r^{\prime\alpha}~,&t^{\prime\alpha} &\mapsto t^\alpha~,~~r^{\prime\alpha}\mapsto r^\alpha~, \\
b^a &\mapsto 0~,~~ h^a\mapsto 0~, &&\\
c^\mu &\mapsto f^\mu_a {\rm cs}^{\prime a}~,&&\\
g^\mu &\mapsto f^\mu_a P^{\prime a}~.&&
\end{aligned}
\end{equation}
Together with the 2-morphism in~\eqref{eq:two_morphism} these again establish the equivalence between $\weil(\frg)$  and $\weil(\twastringsk(\frg))$. 

As $g^\mu$ is now mapped to the invariant polynomial $P^a$, this twisted algebra now allows for the diagram
\begin{equation}\label{eq:g_twisted_g_diagram}
\begin{tikzcd}[arrow style=tikz, >=angle 45]
\ce(\frg) & \ce(\twastringsk)\arrow[l] \\
\weil(\frg)\arrow[u] & \weil(\twastringsk)\arrow[l]\arrow[u] \\
\inv(\frg)\arrow[u] & \inv(\twastringsk)\rlap{~,} \arrow[l]\arrow[u]
\end{tikzcd}
\end{equation}
to be commutative and, therefore, for the lifted connection to be well-defined.

In summary, we have replaced the string algebra $\astringsk(\frg)$ with the algebra $\twastringsk(\frg)$, which corresponds to a twisted Weil algebra, as this allows for a consistent lift of a $\frg$-connection. Even though all of $\weil(\frg)$, $\weil(\astringskext(\frg))$ and $\weil(\twastringsk(\frg))$ are equivalent as $L_\infty$-algebras and we should, in principle, be free to choose any equivalent description, it is only $\weil(\twastringsk(\frg))$ that provides a suitable lift. This is due to the fact that the equivalences given in~\eqref{eq:final_equivalence} additionally preserve the commutativity of the whole sequence in~\eqref{eq:g_twisted_g_diagram}. 

Furthermore, these equivalences mix the components in $\frg^\ast[1]$ and $\frg^\ast[2]$ of $\weil(\twastringsk)$, which leads to modified expressions for the curvatures for the twisted cases. Let us summarize the relevant data for $\twastringsk(\frg)$ in the multi-bracket point of view here. The underlying space is given by
\begin{equation}\label{eq:skeletal_twist_space}
\twastringsk(\frg) = \big(\,\,\frg \overset{0}{\longleftarrow} \FR[1]\overset{{\rm id}}{\longleftarrow}\FR[2]\,\,\big)~,
\end{equation}
in addition to which we have the maps in~\eqref{eq:skel_string_algebra_brackets} and the Killing form
\begin{equation}
\kappa:\frg\otimes\frg\to\FR[1]~,  \kappa(x_1,x_2) =(x_1,x_2)~.
\end{equation}
The curvatures are modified to be
\begin{equation}\label{eq:curvatures_twisted_skel}
\begin{aligned}
\CF &= \dd A + \tfrac12 \mu_2(A,A)~,~~\CG = \dd C~,\\
\CH &= \dd B + \tfrac16 \mu_3(A,A,A) - \kappa(A,\CF)+\mu_1(C)~,\\
\end{aligned}
\end{equation}
with their Bianchi identities given by
\begin{equation}\label{eq:twisted_skeletal_bianchi}
\begin{aligned}
\dd \CF &= -\mu_2(A,\CF)~,~~\dd \CG = 0~,\\
\dd \CH &= -\kappa(\CF,\CF)+\mu_1(\CG)~.\\
\end{aligned}
\end{equation}

The loop model $\astringl(\frg)$ allows for an analogous discussion. We first extend by adding an additional copy of $\FR$ to arrive at
\begin{equation}
\astringl^{\rm ext}(\frg)\,\coloneqq\,\big(\, \astringl(\frg)\overset{\mu_1}{\longleftarrow} \FR[2]\,\big)~,
\end{equation}
where $\mu_1=\text{id}$ is again the only additional structure map. Again, looking at the equivalences explicitly leads us to modify the differential to allow the analogous diagram to commute. In terms of the multi-bracket viewpoint, the role of the cocycle $\mu_3(A,A,A)$ is now played by the cocycle $\mu_2(A,B)$ and the modified differential leads to the introduction of the additional map
\begin{equation}
\begin{aligned}
\kappa: P_0\frg \otimes P_0\frg \to \Omega\frg&\oplus\FR~,\\
\kappa(\gamma_1,\gamma_2) &= \big(\,\chi([\gamma_1,\gamma_2])\, ,\,2\int\limits_0^1 \dd \tau (\dot{\gamma}_1,\gamma_2)\,\big)~,
\end{aligned}
\end{equation}
where $\chi:P_0\frg\to\Omega\frg$ is given by $\chi(\gamma) = \gamma-\partial(\gamma)\cdot \tau$. Here, $\kappa$ is now a more general map playing the role of the Killing form: analogously to the identity $\mu_3 = \kappa\circ\mu_2$, we now have $\mu_2 = \kappa\circ\mu_1$.

We, thus, arrive at the \textit{twisted Weil algebra in the loop model} and the corresponding algebra $\twastringl(\frg)$ with modified curvatures
\begin{equation}\label{eq:curvatures_twisted_loop}
\begin{aligned}
\CF &=\dd A +\tfrac12 \mu_2(A,A) + \mu_1(B)~,~~\CG = \dd C~, \\
\CH &= \dd B + \mu_2(A,B) - \kappa(A,\CF)+ \mu_1(C)~,\\
\end{aligned}
\end{equation}
and their respective Bianchi identities
\begin{equation}
\begin{aligned}
\dd \CF &= -\mu_2(A,\CF) + \mu_1(\kappa(A,\CF))+\mu_1(\CH)~,\\
\dd \CH &= -\kappa(\CF,\CF) + \mu_1(\CG)~,~~\dd \CG = 0~.\\
\end{aligned}
\end{equation}

The modified algebras $\twastringsk(\frg)$ and $\twastringl(\frg)$ are again categorically equivalent with the equivalence exhibited by the same maps as in~\cite{Baez:2005sn}. In~\cite{Saemann:2017rjm}, it was shown that with these twisted versions the fake flatness condition is lifted and one can indeed construct a non-abelian self-dual string. This suggest that the twisted Weil algebras provide interesting gauge structures with relevance to string- and M-theory. 

The above procedure also applies to larger algebras: the gauge structure that was used in~\cite{Saemann:2017zpd} to construct a six-dimensional superconformal model is a larger version of the string algebras discussed here. Finding appropriate twisted curvatures for this enlarged gauge structure would lead to interesting modifications of the model in~\cite{Saemann:2017zpd}, which will be addressed in a future publication.

\bibliographystyle{prop2015}
\bibliography{allbibtex}

\begin{thebibliography}{10}

\bibitem{Saemann:2017rjm}
C.~Saemann and L.~Schmidt,
{\em {The non-Abelian self-dual string and the $(2,0)$-theory},}
{\tt \href{http://www.arxiv.org/abs/1705.02353}{1705.02353 [hep-th]}}.

\bibitem{Saemann:2017zpd}
C.~Saemann and L.~Schmidt,
{\em {Towards an M5-brane model I: a $6d$ superconformal field theory},}
\href{http://dx.doi.org/10.1063/1.5026545}{J. Math. Phys. {\bf 59}  (2018)
  043502} [{\tt \href{http://www.arxiv.org/abs/1712.06623}{1712.06623
  [hep-th]}}].

\bibitem{Sati:2008eg}
H.~Sati, U.~Schreiber, and J.~Stasheff,
{\em $L_\infty$-algebra connections and applications to String- and
  Chern--Simons $n$-transport,}
in: ``Quantum Field Theory,'' eds. B. Fauser, J. Tolksdorf and E. Zeidler, p.
  303, Birkh{\"a}user 2009
[{\tt \href{http://www.arxiv.org/abs/0801.3480}{0801.3480 [math.DG]}}].

\bibitem{Sati:2009ic}
H.~Sati, U.~Schreiber, and J.~Stasheff,
{\em {Differential twisted string and five-brane structures},}
\href{http://dx.doi.org/10.1007/s00220-012-1510-3}{Commun. Math. Phys. {\bf
  315}  (2012) 169} [{\tt \href{http://www.arxiv.org/abs/0910.4001}{0910.4001
  [math.AT]}}].

\bibitem{Baez:2010ya}
J.~C.~Baez and J.~Huerta,
{\em {An invitation to higher gauge theory},}
\href{http://dx.doi.org/10.1007/s10714-010-1070-9}{Gen. Relativ. Gravit. {\bf
  43}  (2011) 2335} [{\tt \href{http://www.arxiv.org/abs/1003.4485}{1003.4485
  [hep-th]}}].

\bibitem{Baez:2002jn}
J.~C.~Baez,
{\em {Higher Yang--Mills theory},}
{\tt \href{http://www.arxiv.org/abs/hep-th/0206130}{hep-th/0206130}}.

\bibitem{Fiorenza:2010mh}
D.~Fiorenza, U.~Schreiber, and J.~Stasheff,
{\em {\v{C}ech cocycles for differential characteristic classes -- an
  infinity-Lie theoretic construction},}
\href{http://dx.doi.org/10.4310/ATMP.2012.v16.n1.a5}{Adv. Th. Math. Phys. {\bf
  16}  (2012) 149} [{\tt \href{http://www.arxiv.org/abs/1011.4735}{1011.4735
  [math.AT]}}].

\bibitem{Waldorf:2009uf}
K.~Waldorf,
{\em String connections and Chern--Simons theory,}
\href{http://dx.doi.org/10.1090/S0002-9947-2013-05816-3}{Trans. Amer. Math.
  Soc. {\bf 365}  (2013) 4393} [{\tt
  \href{http://www.arxiv.org/abs/0906.0117}{0906.0117 [math.DG]}}].

\bibitem{Stolz:2004aa}
S.~Stolz and P.~Teichner,
{\em What is an elliptic object?,}
in ``Topology, geometry and quantum field theory,'' volume~308~of London Math.
  Soc. Lecture Note Ser., pages~247-343. Cambridge Univ. Press, Cambridge,
  2004.

\bibitem{Carey:0410013}
A.~L.~Carey, S.~Johnson, M.~K.~Murray, D.~Stevenson, and B.-L.~Wang,
{\em Bundle gerbes for Chern--Simons and Wess--Zumino--Witten theories,}
\href{http://dx.doi.org/10.1007/s00220-005-1376-8}{Commun. Math. Phys. {\bf
  259}  (2005) 577} [{\tt
  \href{http://www.arxiv.org/abs/math.DG/0410013}{math.DG/0410013}}].

\bibitem{Samtleben:2011fj}
H.~Samtleben, E.~Sezgin, and R.~Wimmer,
{\em {$(1,0)$ superconformal models in six dimensions},}
\href{http://dx.doi.org/10.1007/JHEP12(2011)062}{JHEP {\bf 1112}  (2011) 062}
  [{\tt \href{http://www.arxiv.org/abs/1108.4060}{1108.4060 [hep-th]}}].

\bibitem{Lack:2010sj}
S.~Lack,
{\em {A 2-categories companion},}
\href{http://dx.doi.org/10.1007/978-1-4419-1524-5_4}{IMA Vols. Math. Apps. {\bf
  152}  (2010)~73} [{\tt
  \href{http://www.arxiv.org/abs/math/0702535}{math/0702535}}].

\bibitem{Baez:2003aa}
J.~Baez and A.~S.~Crans,
{\em Higher-dimensional algebra VI: Lie 2-algebras,}
\href{http://tac.mta.ca/tac/volumes/12/15/12-15.pdf}{Theor. Appl. Categor. {\bf
  12}  (2004) 492} [{\tt
  \href{http://www.arxiv.org/abs/math.QA/0307263}{math.QA/0307263}}].

\bibitem{contrib:saemann}
C.~Saemann,
{\em Higher structures, self-dual strings and 6d superconformal field
  theories,}
contribition to this volume.

\bibitem{Lada:1992wc}
T.~Lada and J.~Stasheff,
{\em {Introduction to sh Lie algebras for physicists},}
\href{http://dx.doi.org/10.1007/BF00671791}{Int. J. Theor. Phys. {\bf 32}
  (1993) 1087} [{\tt
  \href{http://www.arxiv.org/abs/hep-th/9209099}{hep-th/9209099}}].

\bibitem{Lada:1994mn}
T.~Lada and M.~Markl,
{\em {Strongly homotopy Lie algebras},}
\href{http://dx.doi.org/10.1080/00927879508825335}{Commun. Alg. {\bf 23}
  (1995) 2147} [{\tt
  \href{http://www.arxiv.org/abs/hep-th/9406095}{hep-th/9406095}}].

\bibitem{0387950680}
Y.~Felix, S.~Halperin, and J.-C.~Thomas,
{\em Rational homotopy theory (Graduate texts in mathematics),} Springer, New
  York, 2001.

\bibitem{Fregier:2013dda}
Y.~Fregier, C.~L.~Rogers, and M.~Zambon,
{\em {Homotopy moment maps},}
\href{http://dx.doi.org/10.1016/j.aim.2016.08.012}{Adv. Math. {\bf 303}  (2016)
  954} [{\tt \href{http://www.arxiv.org/abs/1304.2051}{1304.2051 [math.DG]}}].

\bibitem{Kontsevich:1997vb}
M.~Kontsevich,
{\em {Deformation quantization of Poisson manifolds, I},}
\href{http://dx.doi.org/10.1023/B:MATH.0000027508.00421.bf}{Lett. Math. Phys.
  {\bf 66}  (2003) 157} [{\tt
  \href{http://www.arxiv.org/abs/q-alg/9709040}{q-alg/9709040}}].

\bibitem{Baez:2005sn}
J.~C.~Baez, D.~Stevenson, A.~S.~Crans, and U.~Schreiber,
{\em {From loop groups to 2-groups},}
\href{http://projecteuclid.org/euclid.hha/1201127333}{Homol. Homot. Appl. {\bf
  9}  (2007) 101} [{\tt
  \href{http://www.arxiv.org/abs/math.QA/0504123}{math.QA/0504123}}].

\bibitem{Cartan:1949aaa}
H.~Cartan,
{\em {Cohomologie r{\'e}elle d'un espace fibr{\'e} principal diff{\'e}rentielle
  I, II,},}
{S{\'e}minaire Henri Cartan} {\bf 2}  (1950).

\bibitem{MR0042426}
H.~Cartan,
{\em Notions d'alg\`ebre diff\'{e}rentielle; application aux groupes de {L}ie
  et aux vari\'{e}t\'{e}s o\`u op\`ere un groupe de {L}ie,}
in: Colloque de topologie (espaces fibr\'{e}s), p.15--27, {B}ruxelles, 1950..

\bibitem{Atiyah:1957}
M.~F.~Atiyah,
{\em Complex analytic connections in fibre bundles,}
\href{http://dx.doi.org/10.1090/S0002-9947-1957-0086359-5}{Trans. Amer. Math.
  Soc. {\bf 85}  (1957) 181}.

\bibitem{Bojowald:0406445}
M.~Bojowald, A.~Kotov, and T.~Strobl,
{\em Lie algebroid morphisms, Poisson sigma models, and off-shell closed gauge
  symmetries,}
\href{http://dx.doi.org/10.1016/j.geomphys.2004.11.002}{J. Geom. Phys. {\bf 54}
   (2005) 400} [{\tt
  \href{http://www.arxiv.org/abs/math.DG/0406445}{math.DG/0406445}}].

\bibitem{Kotov:2010wr}
A.~Kotov and T.~Strobl,
{\em {Generalizing geometry - algebroids and sigma models},}
in ``Handbook on Pseudo-Riemannian Geometry and Supersymmetry,'' ed. V. Cortes
[{\tt \href{http://www.arxiv.org/abs/1004.0632}{1004.0632 [hep-th]}}].

\bibitem{Gruetzmann:2014ica}
M.~Gruetzmann and T.~Strobl,
{\em {General Yang--Mills type gauge theories for p-form gauge fields: from
  physics-based ideas to a mathematical framework OR from Bianchi identities to
  twisted Courant algebroids},}
\href{http://dx.doi.org/10.1142/S0219887815500097}{Int. J. Geom. Meth. Mod.
  Phys. {\bf 12}  (2014) 1550009} [{\tt
  \href{http://www.arxiv.org/abs/1407.6759}{1407.6759 [hep-th]}}].

\bibitem{Baez:2004in}
J.~C.~Baez and U.~Schreiber,
{\em Higher gauge theory: 2-connections on 2-bundles,}
{\tt \href{http://www.arxiv.org/abs/hep-th/0412325}{hep-th/0412325}}.

\bibitem{Saemann:2011nb}
C.~Saemann and M.~Wolf,
{\em {On twistors and conformal field theories from six dimensions},}
\href{http://dx.doi.org/10.1063/1.4769410}{J. Math. Phys. {\bf 54}  (2013)
  013507} [{\tt \href{http://www.arxiv.org/abs/1111.2539}{1111.2539
  [hep-th]}}].

\bibitem{Saemann:2012uq}
C.~Saemann and M.~Wolf,
{\em {Non-Abelian tensor multiplet equations from twistor space},}
\href{http://dx.doi.org/10.1007/s00220-014-2022-0}{Commun. Math. Phys. {\bf
  328}  (2014) 527} [{\tt \href{http://www.arxiv.org/abs/1205.3108}{1205.3108
  [hep-th]}}].

\end{thebibliography}
\end{document}